\documentclass{article}
\usepackage{dcase2020,amsmath,graphicx,url,times,booktabs, tabularx}
\usepackage{tablefootnote}
\usepackage{comment}
\usepackage{mathtools}
\usepackage{amssymb}
\usepackage{csquotes}
\usepackage{multirow}
\usepackage{hyperref}
\usepackage{comment}
\usepackage{float}
\usepackage{enumitem}
 \usepackage[table,xcdraw]{xcolor}

\hypersetup{
    colorlinks=true,
    linkcolor=blue,
    filecolor=magenta,      
    urlcolor=cyan,
}

\title{Low-complexity CNNs for acoustic scene classification}

\name{Arshdeep Singh, James A King, Xubo Liu, Wenwu Wang, Mark D. Plumbley}
\address{Centre for Vision, Speech and Signal Processing (CVSSP) \\ University of Surrey, UK\\
Email: $\{arshdeep.singh,james.a.king,xubo.liu,w.wang,m.plumbley\}$@surrey.ac.uk}

\begin{document}

\ninept
\maketitle

\begin{sloppy}
\begin{abstract}
    
This technical report describes the SurreyAudioTeam22’s submission for DCASE 2022 ASC Task 1, Low-Complexity Acoustic Scene Classification (ASC). 
The task has two rules, (a) the  ASC framework should have maximum 128K parameters, and (b) there should be a maximum of 30 millions multiply-accumulate operations (MACs) per inference. In this report, we present  low-complexity systems for ASC that follow the rules intended for the task.

\end{abstract}


\section{Introduction}
\label{sec:intro}

Convolutional neural networks (CNNs) have shown  state-of-the-art performance  in comparison to traditional hand-crafted methods in various domains \cite{gu2018recent}.  However, CNNs are resource hungry due to their large size and  memory \cite{simonyan2014very, krizhevsky2012imagenet}, and hence it is difficult to deploy CNNs on resource constrained devices. For example, Cortex-M4 devices (STM32L496@80MHz or Arduino Nano 33@64MHz) have a  maximum allowed limit is 128K parameters and 30M multiply-accumulate operations (MACs) per inference corresponding to an audio of 1 second length. Thus, the issue of reducing the size  and the computational cost of CNNs has drawn a significant amount of attention in the detection and classification of acoustic scenes and events (DCASE) research community.

This report aims to design \enquote{low-complexity CNNs} for ASC which have a maximum number of parameters less than 128K and a maximum number of MACs less than 30M. The baseline CNN designed for the DCASE 2022 task 1 has 46512 parameters, 29.24M MACs and the accuracy is approximately 43\% with 1.575 log-loss.


We propose the following steps to obtain low-complexity CNNs.
\begin{enumerate}[label=(\alph*)]
    \item We design a \enquote{low-complexity} CNN that has a fewer parameters and MACs with  better performance  than that of the baseline CNN. 
    \item A filter pruning method is applied to compress the \enquote{low-complexity CNN} of (a) further. Subsequently, we quantize  each parameter from float32 to INT8 data type, reducing networks memory by 4x.
    \item An ensemble approach is proposed which combines predictions obtained from (b) low-complexity CNNs.
\end{enumerate}



The rest of the report is organized as follows. 
In section \ref{sec: low-complexity CNNs}, a procedure to obtain low-complexity CNN and an ensemble framework  is described.  A brief overview of dataset used, features used for experimentation and experimental setup are described in Section \ref{sec: dataset used}. Section \ref{sec: perfromance}  presents  experimental analysis. The submitted entries for the DCASE 2022 task 1 challenge are included in \ref{sec: submitted entried}. Finally, conclusion is presented in Section \ref{sec: conclusion}.

\section{Low-complexity CNNs}
\label{sec: low-complexity CNNs}
\noindent \textbf{Proposed low-complexity architecture:} We propose a CNN architecture which  consists of three convolutional layers (C1, C2 \& C3), two pooling layers (P1, P2), a dense layer (D1) and a classification layer. The details of different layers is given in Table \ref{tab: low-complxity architecutre}. The proposed architecture requires approximately 5M MACs to produce an output corresponding to an input of size  (40 x 51), and has 14886 parameters.

\noindent \textbf{Filter Pruning:} To eliminate redundant filters from the low-complexity architecture as given in Table \ref{tab: low-complxity architecutre}, we apply a filter pruning strategy. For each convolutional layer, we identify filter pairs which are similar. Our hypothesis is that similar filters produce similar output and hence, contribute to redundancy only. Therefore, one of the similar filters can be eliminated. The similarity between the filters is measured using a cosine distance. We identify the closest filter pairs for each layer separately. A filter from each pair is deemed redundant and eliminated from the network. More information about the similarity based  filter pruning method can be found at  \cite{singh2022passive}.  

The number of redundant filters obtained after performing  similarity-based filter pruning for C1 layer is 4 out of 16, C2 layer is 4 out of 16 and C3 layer is 10 out of 32. We obtain 6 different pruned networks that are obtained after pruning C1 layer only, C2 layer only, C3 layer only, C1 and C2 layers, C2 and C3 layers, C1 and C2 and C3 layers. The number of MACs and the number of parameters for each pruned network are given in Table \ref{tab:  various CNNs DCASE2022}.

\noindent \textbf{Quantization:} To reduce the network size further, we  perform quantization on parameters of each pruned network using TFlite optimization, and quantized the network parameters from 32-bit floating point to 8-bit integers (INT8). 


\noindent \textbf{An ensemble framework:} Next,  the predictions obtained from various pruned networks are aggregated together in an ensemble framework. The total number of parameters in the ensemble framework that aggregates predictions from all 6 pruned networks are 70.97K and the total number of MACs are 23.84M.

\begin{table*}[h]
\caption{Proposed low-complexity architecture. Here, BN stands for batchnormalization, tanh is hyperbolic tangent activation function and ReLU is a rectified linear unit activation function.}
\vspace{0.25cm}
\centering
\label{tab: low-complxity architecutre}
\resizebox{0.75\textwidth}{!}{
\begin{tabular}{ccccccc}
\toprule
Layer name & Description                    & Number of filters/dense units & Filter/pooling size & input     & output    \\		\midrule
{C1} & {Convolution + BN + tanh}  & {16}                            & {(3 x 3)}             & {(40 x 51)} & {(40 x 51)} \\ \midrule
{C2} & {Convolution + BN + ReLU}  & {16 }                           & {(3 x 3)}             & {(40 x 51)} & {(40 x 51)} \\ \midrule
{P1} & {Average Pooling}          & {- }                         & {(5 x 5)}              & {(40 x 51)}  & {(8 x 10)}   \\\midrule
{C3} & {Convolution + BN + tanh}  & {32}                            & {(3 x 3)}             & {(8 x 10)}  & {(8 x 10)}  \\\midrule
{P2} & {Average Pooling}          & {-}                            & {(4 x 10)}             & {(8 x 10)}  & {(2 x 1)}   \\ \midrule
{Dense} & {Dense + tanh}             & {100}                           & {1}                   & {64}        & {100}       \\ 		\midrule
{Classification} & {Classification + softmax} & {10}                  &   { 1 }                & {100}       & {10}              \\    \bottomrule
\end{tabular}
}
\end{table*}

\begin{table*}[h]
	\caption{Various low-complexity CNNs obtained after pruning and applying quantization (INT8). Here, the frameworks in the bold entries are used for evaluation on DCASE challenge dataset.}
	\centering
	\label{tab:  various CNNs DCASE2022}
	\vspace{0.25cm}
	\Large
	\resizebox{0.85\textwidth}{!}{
	\begin{tabular}{ccccccc} \toprule
		{Sr No.}& {Network Name} & {Pruned layer}  & {Architecture (C1-C2-C3-Dense)} & {Parameters} & {Size (KB)} & {MACs (millions)}  \\
		\midrule
		\textbf{1} &{\textbf{Unpruned low-complexity}} & {\textbf{NA}} &{\textbf{16-16-32-100}}& {\textbf{14886}} & { \textbf{18.59} } & {\textbf{5.41}}  \\
		\midrule
		{2}& {Pruned\_C1} &{C1} & {12-16-32-100} &{14254} & { 17.86 } & {4.16}  \\
		\midrule
		{\textbf{3}}&{\textbf{Pruned\_C2}} & {\textbf{C2}} & {\textbf{16-12-32-100}} & {\textbf{13138}} & { \textbf{16.85}} & {\textbf{4.13}} \\
		\midrule
		{4} & {Pruned\_C3} &{C3} & {16-16-22-100} & {11396} & { 15.11 } & {5.29} \\
        \midrule
		{5} & {Pruned\_C12} &{C1 + C2} & {12-12-32-100} & {12650} & { 16.26 } & {3.18} \\
		\midrule
		{6}&{Pruned\_C23} &{C2 + C3} & {16-12-22-100} &{10008} &  { 13.73} & {4.04}  \\
		\midrule
		{7}& {Pruned\_C123} &{C1 + C2 + C3} & {12-12-22-100} & {9520} & { 13.14 } & {3.08} \\
	    \bottomrule
	\end{tabular}}
\end{table*}

\begin{table*}[t]
	\caption{Performance obtained when different pruned and quantized networks are used in the ensemble framework. Here, the frameworks in the bold entries are used for evaluation on DCASE challenge dataset.}
	\Large
	\centering
	\label{tab:  ensemble DCASE2022}
	\vspace{0.25cm}
	\resizebox{0.9\textwidth}{!}{
	\begin{tabular}{cccccccc} \toprule
		{Sr No.}& {Ensemble framework} & {Number of parameters}  & {MACs (millions)} & {Size (KB)} & {Accuracy (\%)} & {Log-loss}  \\
		\midrule
		{1}& {All pruned networks except Pruned\_C1} &{56712} & {19.72} & {75.09 } & {47.14} & {1.394}  \\
		\midrule
		{2}&{All pruned networks except Pruned\_C2} & {57828} & {19.75} & {76.10} & {47.10} & {1.396} \\
		\midrule
		{\textbf{3}} & {\textbf{All pruned networks except Pruned\_C3}} &{\textbf{59570}} & {\textbf{18.60}} & {\textbf{77.84}} & {\textbf{47.26}} & {\textbf{1.392}} \\
        \midrule
		{4} & {All pruned networks except Pruned\_C12} &{58316} & {20.70} & {76.69} & {47.45} & {1.394} \\
		\midrule
		{\textbf{5}}&{\textbf{All pruned networks except Pruned\_C23}} &{\textbf{60958}} & {\textbf{19.84}} & {\textbf{79.22}} & {\textbf{47.52}} & {\textbf{1.389}}  \\
		\midrule
		{6} & {All pruned networks except Pruned\_C123} &{61446} & {20.80} & {79.81} & {47.35} & {1.392} \\
				\midrule
		{7} & {Ensemble on all pruned networks} & {70966} & {23.84} & {92.95} & {47.45} & {1.389} \\
       \bottomrule
	\end{tabular}}
\end{table*}

\section{Experimental dataset, feature extraction and experimental setup}
\label{sec: dataset used}
\noindent \textbf{Dataset used:} We use the TAU Urban Acoustic Scenes 2022 Mobile, development dataset \cite{Heittola2020}. The dataset contains recordings from 12 European cities in 10 different acoustic scenes using 4 different devices. Each audio recording has 1 second length. The dataset is divided in training and validation sets. The training dataset consists of  139620 audio examples and the validation dataset consists of 29680 audio examples.

\noindent \textbf{Feature extraction:} For time-frequency representations, log-mel band energies of size (40 $\times$ 51) corresponding to an audio signal of 1 second length are extracted. A Hamming asymmetric window of length  40ms, and a hop length of 20ms is used to extract magnitude spectrogram. Next, log-mel spectrogram is computed using 40 mel bands.

\noindent \textbf{Experimental Setup:} The unpruned low-complexity network is trained using the training dataset with a batch size of 64 with an Adam optimizer for 1000 epochs. A categorical cross-entropy loss function is used during the training process. We apply an early stopping criterion to yield the best network that gives the minimum log-loss for the validation dataset.

After obtaining the trained  unpruned low-complexity CNN, the filter pruning strategy described in Section \ref{sec: low-complexity CNNs} is applied to eliminate redundant filters. To regain the loss in performance due to pruning, the pruned networks are fine-tuned with a similar process to that used for the training of the unpruned low-complexity network.

\section{Performance Analysis}

\label{sec: perfromance}
Figure \ref{fig: performance of the pruned network} shows the accuracy and the log-loss obtained for the unpruned low-complexity network and the various pruned networks. The unpruned low-complexity network gives 1.475 log-loss and 45.92\% accuracy.  Eliminating filters from the unpruned  network 
results in a significant reduction in performance, but this is almost entirely restored after fine-tuning.

The performance obtained after aggregating predictions from various pruned networks is given in Table \ref{tab:  ensemble DCASE2022}. The ensemble framework improves the performance in comparison to that of individual pruned networks.

\begin{figure}[H]
    \centering
    \includegraphics[scale=0.41]{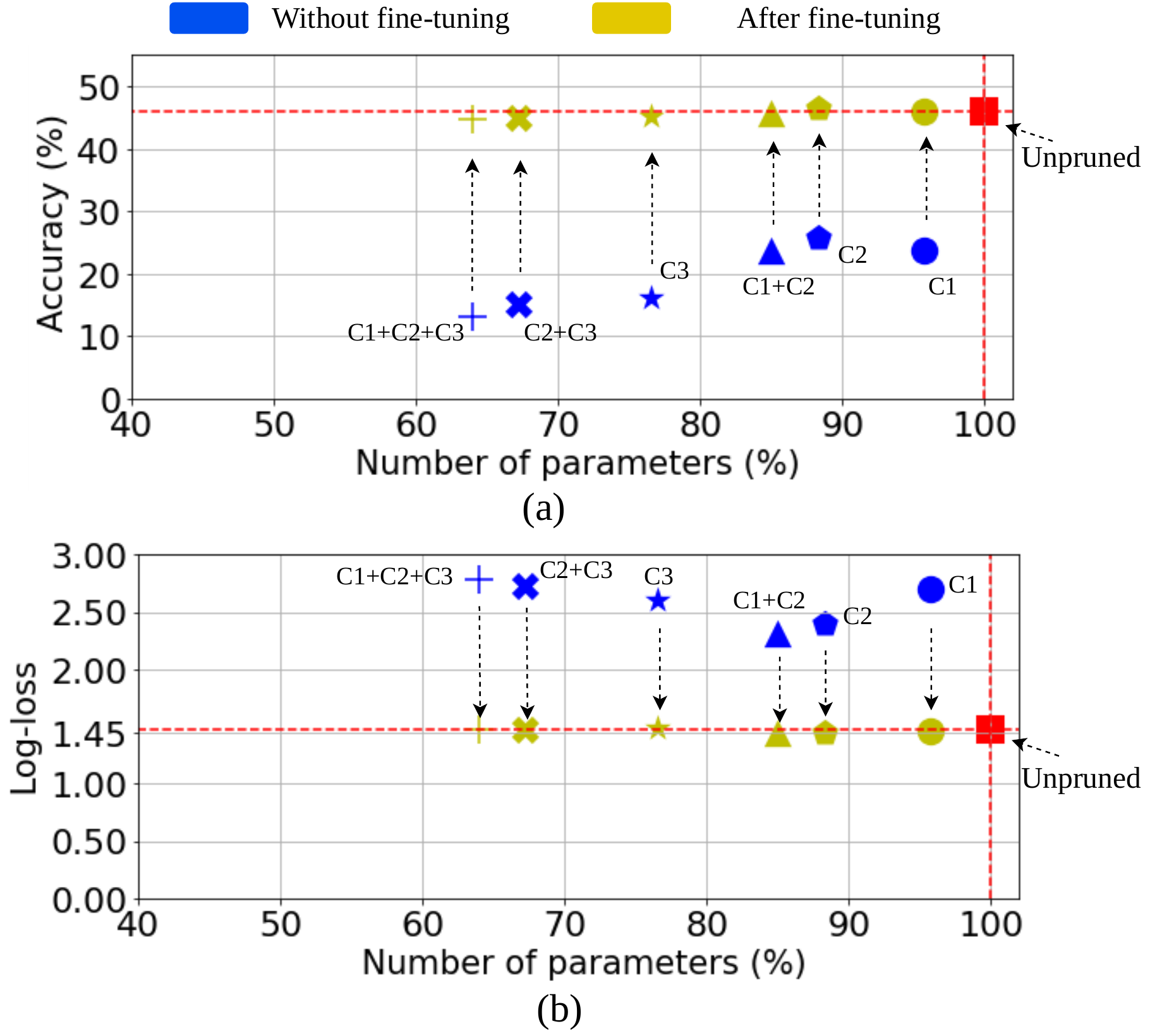}
    \caption{(a) Accuracy and (b) log-loss obtained after pruning different intermediate layers (C1, C2, C3, C1+C2, C2+C3, C1+C2+C3) in the unpruned low-complexity CNN. The accuracy and the log-loss is obtained with and without performing the fine-tuning of the pruned network.}
    \label{fig: performance of the pruned network}
\end{figure}

\noindent \textbf{Comparison with DCASE 2022 Task 1 baseline network:} In comparison to the DCASE 2022 task 1 baseline network, the proposed unpruned low-complexity network improves the accuracy by approximately 3 percentage points, improves log-loss by 0.102 points  with  approximately 5x reduction in MACs and 3x reduction in the parameters. Utilizing ensemble of pruned networks improves the accuracy by approximately 4.5 percentage points, improves log-loss by 0.19 points with 1.5x reduction in MACs, however requires approximately 14K more parameters in comparison to that of the DCASE 2022 task 1 baseline network.

\begin{table}[h]
	\caption{Performance obtained when different pruned networks are used in the ensemble framework.}
	\Large
	\centering
	\label{tab: performance comparison}
	\vspace{0.25cm}
	\resizebox{0.5\textwidth}{!}{
	\begin{tabular}{ccccc} \toprule
		{Framework} & {Accuracy} & {log-loss} &{ MACs (million)}  & {Parameters}  \\ 
		\midrule
		{DCASE 2022  Task 1 baseline } & {43\%} &{1.575} & {29.23} &{46512}  \\
		\midrule
		{Unpruned low-complexity CNN} & {45.92\%} & {1.475} & {5.44} & {14886}\\
		\midrule
		{Ensemble on all pruned networks  except Pruned\_C23} & {47.52\%} & {1.389} & {19.84} & {60958} \\
       \bottomrule
	\end{tabular}}
\end{table}

\section{Submitted Entries}
\label{sec: submitted entried}
In this section,  a detail of various submitted models is described.

\begin{enumerate}

    \item \textbf{Singh\_Surrey\_task1\_1 (SurreyAudioTeam22\_1, Surrey\_4M):} This  challenge entry includes predictions from the Pruned\_C2 network. Please see Table \ref{tab:  various CNNs DCASE2022}, Sr No. 3, for more detail of the network.
    \item \textbf{Singh\_Surrey\_task1\_2 (SurreyAudioTeam22\_2, Surrey\_5M):}  This  challenge entry includes predictions from the unpruned low-complexity network. Please see Table \ref{tab:  various CNNs DCASE2022}, Sr No. 1, for more detail of the network.
    \item \textbf{Singh\_Surrey\_task1\_3 (SurreyAudioTeam22\_3, Surrey\_19M):} This challenge entry includes predictions from the ensemble framework which combines all pruned networks except Pruned\_C3. Please see Table \ref{tab:  ensemble DCASE2022}, Sr No. 3, for more detail of the network.
     \item \textbf{Singh\_Surrey\_task1\_4 (SurreyAudioTeam22\_4, Surrey\_20M):} This challenge entry includes predictions from the ensemble framework which combines all pruned networks except Pruned\_C23. Please see Table \ref{tab:  ensemble DCASE2022}, Sr No. 5, for more detail of the network.
\end{enumerate}

The pruned quantized networks can be found at this link \footnote{\href{https://github.com/Arshdeep-Singh-Boparai/DCASE2022_Challenge.git}{Link: Pruned Quantized models, evaluation scripts.}}. The network size, the number of MACs are computed using this link\footnote{\href{https://github.com/AlbertoAncilotto/NeSsi}{Link: Model size and complexity calculation.}}.

\section{Conclusion}
\label{sec: conclusion}
This report focuses on designing low-complexity system for acoustic scene classification. A filter pruning, quantization and ensemble procedure is applied to obtain compressed, accelerated, and low-size CNN. The proposed framework shows promising results in terms of reduction in  parameters and performance.

\section{Acknowledgements}
This work was partly supported by Engineering and Physical Sciences Research Council (EPSRC) Grant EP/T019751/1 \enquote{AI for Sound (AI4S)}, a PhD scholarship from the Centre for Vision, Speech and Signal Processing (CVSSP), Faculty of Engineering and Physical Science (FEPS), University of Surrey, a Research Scholarship from the China Scholarship Council (CSC) No. 202006470010 and a Newton Institutional Links Award from the British Council (Grant number 623805725). For the purpose of open access, the authors have applied a Creative Commons Attribution (CC BY) licence to any Author Accepted Manuscript version arising.
For more information about the AI4S project, please follow the link: \url{https://ai4s.surrey.ac.uk/index}.

\bibliographystyle{IEEEtran}
\bibliography{refs}

%
%
%
%
%
%
%
%
%

\end{sloppy}
\end{document}